%% file: exact-collider-letter-v4.tex
\documentclass[aps,twocolumn,nofootinbib,groupedaddress]{revtex4-1}

\usepackage[english]{babel}
\usepackage{graphicx}
\usepackage{amsmath,amssymb,amsbsy,amstext, amsthm, simplewick, amsfonts}
\usepackage{dcolumn}

\usepackage{subfiles}

\usepackage{hyperref}
\hypersetup{
    colorlinks=true,
    linkcolor=purple,
    filecolor=purple,
    urlcolor=purple,
    citecolor=purple
    }

\usepackage{comment}
\usepackage{bm}
\usepackage{xcolor}
\usepackage{braket}
\usepackage{tensor}
\usepackage{soul}
\usepackage{tikz}
\definecolor{pyblue}{RGB}{31, 119, 180}
\definecolor{pyorange}{RGB}{255, 127, 14}
\definecolor{pygreen}{RGB}{44, 160, 44}
\definecolor{pyred}{RGB}{214, 39, 40}
\usetikzlibrary{intersections, calc, arrows.meta}
\usetikzlibrary{patterns}

\newcommand{\dd}{\mathrm{d}}

\def\mpl{M_{\rm pl}}

\usepackage{colortbl}

\renewcommand{\O}{\mathcal{O}}

\newcommand{\pic}{{\pi_{\rm c}}}
\newcommand{\la}{\lambda}
\newcommand{\n}{{\nu_{\rm eff}}}
\newcommand{\muf}{{\mu_{\rm eff}}}
\newcommand{\meff}{{m_{\rm eff}}}
\newcommand{\ka}{\kappa}
\newcommand{\Wn}{\mathsf{W}}
\newcommand{\Ma}{M}
\newcommand{\Real}{\mathrm{Re}}

\begin{document}

\title{New exact bispectrum shapes in multifield inflation}

\author{Lucas Pinol}
\email{lucas.pinol@phys.ens.fr}

\affiliation{Laboratoire de Physique de l'{\'E}cole Normale Sup{\'e}rieure, ENS, CNRS, Universit{\'e} PSL,\\ Sorbonne Universit{\'e}, Universit{\'e} Paris Cit{\'e}, F-75005, Paris, France}

\begin{abstract}
Using the effective field theory of multiple inflationary fluctuations, we present the first analytical calculation of the primordial bispectrum in which the quadratic mixing between curvature and isocurvature fluctuations is treated non-perturbatively.
Building upon the operator representation of the exact linear solutions proposed in Ref.~\cite{Huenupi:2026abj}, we derive a simpler integral representation for these mixed mode functions.
We prove that all scale-invariant tree-level bispectra reduce to a single vertex diagram, which can be evaluated with a Schwinger-parameter integral over independent pre-computable leg kernels.
We showcase the power of our approach by considering the cubic time-derivative interaction $\dot{\pic}^3$, which leads to a purely single-field, equilateral phenomenology at small mixing.
On the contrary, at strong mixing the obtained bispectrum shapes decorrelate from the equilateral template and become genuinely multifield, with a large amplitude, motivating a dedicated data analysis.
The squeezed limit is obtained analytically in a closed form at any dimensionless mixing strength $\la$ for an isocurvature field of bare mass $m$ and features a cosmological collider signal set by $\nu_{\rm eff} = i \muf=\sqrt{9/4-m^2/H^2-\la^2}$, with an effective mass dressed by $\la$, as previously evidenced in numerical or semi-analytical calculations.
Our results encompass the $\la \ll 1$ limit of usual perturbative calculations, where the amplitude of the signal is necessarily small, but they also surpass them, thus opening a new analytical window into large multifield primordial non-Gaussianities.
\end{abstract}

\maketitle

\paragraph*{\bf Introduction.}

\vspace{0.2cm}
While cosmic inflation is the leading paradigm for our early universe history, its microphysical details remain unknown so far.
Primordial non-Gaussianities are arguably the finest probe of the interactions and field content of inflation.
A massive fluctuation with $m\sim H$ coupled to the curvature perturbation imprints a characteristic non-analytical signal on the soft limits of $n > 2$-point functions: the cosmological collider signal~\cite{Chen:2009we,Chen:2009zp,Noumi:2012vr,Arkani-Hamed:2015bza,Lee:2016vti,Arkani-Hamed:2018kmz}. In the minimal realisation, ubiquitous in multifield scenarios, a massive isocurvature scalar $\sigma$ couples to the Goldstone boson of broken time translations $\pic$ through the dimension-three quadratic mixing $\rho\,\dot{\pic}\,\sigma$ generated by a turning background trajectory~\cite{Chen:2009zp,Achucarro:2010jv}.

Analytical treatments of this system have remained, almost exclusively, perturbative in the dimensionless mixing strength $\la\equiv\rho/H$~\cite{Chen:2009zp,Chen:2017ryl,Arkani-Hamed:2018kmz,Qin:2023ejc,Pinol:2021aun,Aoki:2024uyi}. This expansion is adequate for $\la\ll1$, but this is precisely the regime where the collider signal is small. The regime of largest, potentially observable signal, $\la\gtrsim1$, has so far been accessible only through numerical methods~\cite{An:2017hlx,Werth:2023pfl,Jazayeri:2023xcj,Pinol:2023oux}.

\begin{figure*}[t]\centering
\includegraphics[width=0.46\textwidth]{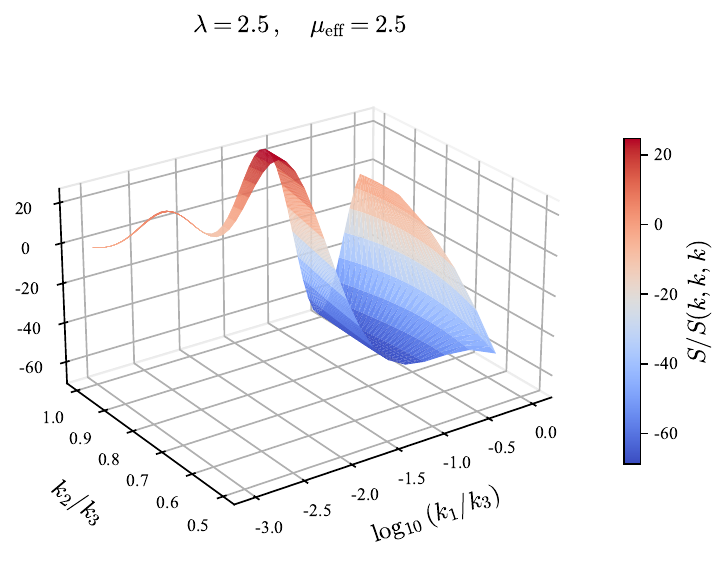}\hfill
\includegraphics[width=0.44\textwidth]{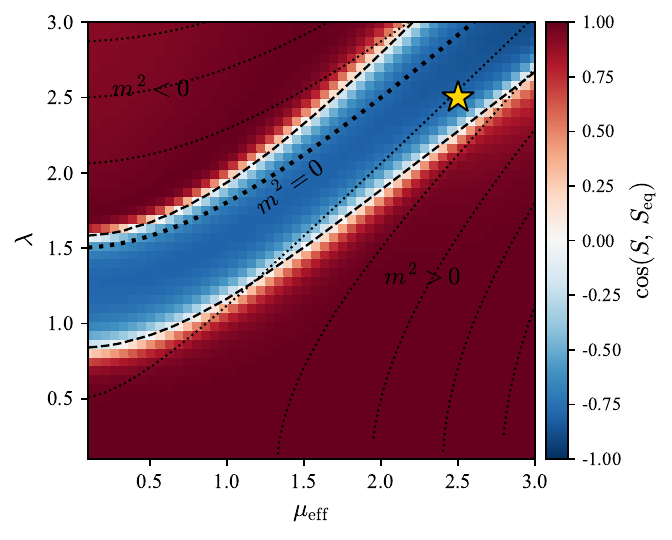}
\caption{\emph{Left:} contribution of the contact $\dot\pic^3$ diagram to the bispectrum shape $S(k_1,k_2,k_3)$, normalized to unity at the equilateral configuration, for $(\la,\muf)=(2.5,2.5)$.
Collider oscillations of frequency $\muf$ invade the mildly squeezed configurations with an amplitude parametrically enhanced compared to the equilateral value (which is suppressed for this parameter choice).
\emph{Right:} cosine correlation~\eqref{eq:cosine} of the exact shape~\eqref{eq:Spi3} with the equilateral template across the $(\la,\muf)$ plane. The star represents the parameters' value of the left panel; dashed lines are $\cos=0$; the dotted lines represent constant bare masses separated by $\Delta( m^2/H^2)=2$ with the $m^2=0$ highlighted, separating the massive from the tachyonic bare theory.
The perturbative regime is dominated by equilateral configurations, the shape fully decorrelates in a band running below the $m^2=0$ line, and equilateral correlation is only partially restored, then degrading, at very strong mixing.}
\label{fig:shape_cosmap}\end{figure*}

\begin{figure*}[t]\centering
\raisebox{-0.5\height}{\includegraphics[width=0.43\textwidth]{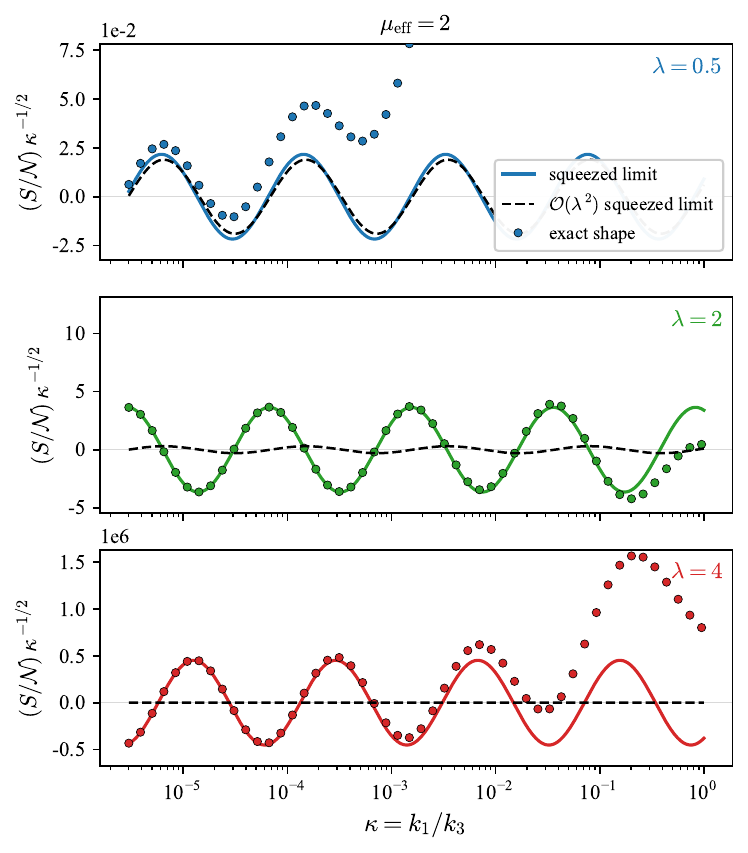}}\hfill
\raisebox{-0.5\height}{\includegraphics[width=0.45\textwidth]{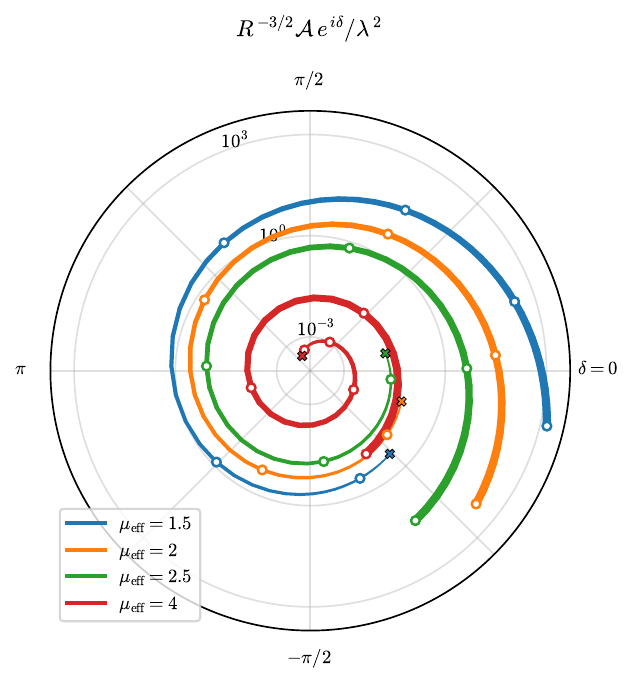}}
\caption{\emph{Left:} cosmological collider signal on isosceles triangles ($k_2=k_3$, $\ka=k_1/k_3$) at $\muf=2$, in units of $\mathcal{N}$ and rescaled by $\kappa^{-1/2}$, for $\la=0.5,2,4$.
Discrete points are drawn from the exact shape~\eqref{eq:Spi3}, solid lines follow from the closed-form squeezed limit~\eqref{eq:clock}; dashed lines are their perturbative, $\la\ll 1$ limit, Eq.~\eqref{eq:Cweak}. 
\emph{Right:} physical complex clock amplitude rescaled by $\lambda^{-2}$,  $R^{-3/2}\mathcal{A}e^{i\delta}/\la^2$, as parametric curves of $\la \in [0.05,6]$ (thickness increases with $\la$) for $\muf=1.5,2,2.5,4$, computed from the squeezed limit~\eqref{eq:clock}.
The crosses denote the prediction from the $\mathcal{O}(\lambda^2)$ expansion in Eq.~\eqref{eq:Cweak} and the six dots denote $\lambda \in \{1, \ldots ,6\}$.}
\label{fig:clock}\end{figure*}

In this \emph{Letter}, and in a companion paper~\cite{TBP} where the diagrammatics is developed in full, we open an analytical window into the strong-mixing regime. Building upon the exact solutions of the coupled linear system recently obtained in Ref.~\cite{Huenupi:2026abj}, which we recast as single integral representations, we prove that every tree-level bispectrum diagram reduces to a one-dimensional Schwinger-parameter integral over independent, pre-computable leg kernels, at any value of $\la$. We showcase the method on the simplest diagram, the ``no-exchange'' bispectrum generated by the self-interaction $\dot{\pi}^3$, deferring the exchange diagrams involving $\sigma$ legs to the companion paper. Despite its single-field appearance, this diagram is dressed by the mixing at all orders and displays a rich multifield phenomenology: while it reduces to the standard equilateral shape in the weak mixing limit $\la\ll 1$, at strong mixing it decorrelates from the equilateral template (Fig.~\ref{fig:shape_cosmap}) as an exponentially enhanced cosmological collider signal, which we obtain in closed form, invades mildly squeezed configurations (Fig.~\ref{fig:clock}).

\vskip 4pt
\paragraph*{\bf Effective field theory.}
We consider the effective field theory (EFT) of inflationary fluctuations~\cite{Creminelli:2006xe,Cheung:2007st} extended by a massive scalar $\sigma$~\cite{Senatore:2010wk, Noumi:2012vr, Pinol:2023oux, Pinol:2024arz}. In unitary gauge, the operators compatible with the residual symmetries are built from $\delta g^{00}=g^{00}+1$ and, at the orders relevant here,
\begin{equation}
\label{eq:SEFT}
\begin{aligned}
   S=\int \dd^4x \sqrt{-g}\,\Big[&\tfrac{1}{2}\mpl^2R+\mpl^2\dot{H}g^{00}-\mpl^2(3H^2+\dot{H})\\
   &+\tfrac{1}{2!}M_2^4(\delta g^{00})^2+\tfrac{1}{3!}M_3^4(\delta g^{00})^3\\
   &-\tfrac{1}{2}(\partial_\mu\sigma)^2-\tfrac{1}{2}m^2\sigma^2+\tilde{M}^3\delta g^{00}\sigma\Big]\,,
\end{aligned}
\end{equation}
with $M_2,M_3,\tilde{M}$ free constant---we restrict ourselves to scale-invariant scenarios in this work---Wilson coefficients and $\sigma$ a pure scalar with unit sound speed. Introducing the Goldstone boson $\pi$ via $t\to t+\pi$, one has $\delta g^{00}=-2\dot{\pi}-\dot{\pi}^2+a^{-2}(\partial_i\pi)^2$ in the decoupling limit, and the curvature perturbation reads $\zeta=-H\pi$. Expanding Eq.~\eqref{eq:SEFT} at quadratic order in the fluctuations gives, at leading order in slow roll,
\begin{equation}
\label{eq:Lquad}
\begin{aligned}
   \frac{\mathcal{L}^{(2)}}{a^3} = &\,\frac{f_\pi^4}{2}\left[\frac{\dot{\pi}^2}{c_s^2}-\frac{(\partial_i \pi)^2}{a^2}\right]-2\tilde{M}^3\dot{\pi}\,\sigma\\
   &+\frac{1}{2}\left[\dot{\sigma}^2- \frac{(\partial_i \sigma)^2}{a^2} - m^2 \sigma^2\right]  \,,
\end{aligned}
\end{equation}
with $f_\pi^4\equiv2\epsilon H^2\mpl^2$ and the speed of sound of propagation defined by $c_s^{-2}\equiv1+4M_2^4/f_\pi^4$. At cubic order, the very same operators generate
\begin{equation}
\label{eq:Lcubic}
\begin{aligned}
   \frac{\mathcal{L}^{(3)}}{a^3} =&\left(2M_2^4-\tfrac43 M_3^4\right)\dot{\pi}^3-2M_2^4\,\dot{\pi}\frac{(\partial_i\pi)^2}{a^2}\\
   &-\tilde{M}^3\sigma\left[\dot{\pi}^2-\frac{(\partial_i \pi)^2}{a^2}\right]\,,
\end{aligned}
\end{equation}
whose sizes are tied to the quadratic theory~\eqref{eq:Lquad} by the non-linearly realised boosts: the $\dot{\pi}(\partial_i\pi)^2$ coupling is fixed by $c_s$, and the $\sigma\pi^2$ interactions by the mixing $\tilde{M}^3$ itself.

For simplicity, we now set $M_2=0$, i.e.~a unit speed of sound of propagation, which also imposes the cancellation of the $\dot{\pi}(\partial_i\pi)^2$ cubic interaction via the non-linearly realised symmetries. Moreover, we focus in this \emph{Letter} on the bispectrum generated by the $\dot{\pi}^3$ operator, which constitutes an independent, zero-exchange diagram, and we omit the cubic interactions of the form $\sigma\pi^2$, an unavoidable Lorentz-invariant combination whose size is fixed by the mixing itself~\cite{Pinol:2023oux} and which generates a distinct single-exchange diagram (see \emph{Regime of validity} below). Canonically normalising $\pic=f_\pi^2\pi$, the quadratic mixing reads $\rho\,\dot{\pic}\,\sigma$ with $\rho=-2\tilde{M}^3/f_\pi^2$ and $\la\equiv\rho/H$, and the retained cubic interaction is $-\lambda_2\dot{\pic}^3$ with $\lambda_2=4M_3^4/(3f_\pi^6)$.

We can then directly give the Hamiltonian densities of the theory. The Legendre transform of $\mathcal{L}^{(2)}+\mathcal{L}^{(3)}$ gives the (non-linear) momenta $p_\pi=a^3(\dot{\pic}+\rho\sigma- 3 \lambda_2 \dot{\pic}^2)$ and $p_\sigma=a^3\dot{\sigma}$.
We choose the free Hamiltonian density to contain the quadratic mixing instead of treating it perturbatively, resulting in
\begin{equation}
\label{eq:H2}
   \mathcal{H}_{\rm free}=\frac{p_\pi^2}{2a^3}+\frac{p_\sigma^2}{2a^3}+\frac{a}{2}(\partial_i\pic)^2 + \frac{a}{2}(\partial_i\sigma)^2
   +\frac{a^3}{2} \meff^2\sigma^2- \rho\,\sigma p_\pi\,,
\end{equation}
where the isocurvature mass is now the \emph{effective} mass
\begin{equation}
\label{eq:meff}
   \meff^2=m^2+\rho^2\,,\qquad \n\equiv\sqrt{9/4-\meff^2/H^2}\,,
\end{equation}
with $\n=i\muf$, $\muf>0$, for a heavy field $\meff>\tfrac32 H$: the bare mass never appears in isolation, and all linear mode functions depend on $(m,\rho)$ only through $(\n,\la)$. This observation analytically explains the effective collider frequency $\muf=\sqrt{\la^2+m^2/H^2-9/4}$ observed numerically and semi-analytically in Refs.~\cite{An:2017hlx,Werth:2023pfl,Pinol:2023oux}. Note also that for a non-tachyonic bare mass, $m^2\geqslant 0$, the field is necessarily effectively heavy at strong mixing, so that collider oscillations are unavoidable for $\lambda \geqslant 3/2$; conversely, at fixed $\muf$, mixing strengths $\la^2>9/4+\muf^2$ correspond to a tachyonic bare mass, to which we come back in the conclusion.
In the interaction picture, our Hamiltonian density is simply given by
\begin{equation}
\label{eq:H3}
   \mathcal{H}_{\rm int}=\lambda_2\, a^3 \big(\dot{\pi}_{\rm c}^{I}\big)^3\,,
\end{equation}
where $\dot{\pi}_{\rm c}^{I}=p^I_\pi/a^3-\rho\sigma^I$ is the interaction-picture velocity, whose mode functions are those of the free theory~\eqref{eq:H2}: despite the derivative mixing, the apparent extra vertices generated by the Legendre transform cancel at cubic order, consistently with the general in-in analysis of Ref.~\cite{Chen:2017ryl}, and first deviations from $\mathcal{H}_{\rm int}=-\mathcal{L}_{\rm int}$ arise at quartic order only.
We omit the superscript $I$ in the following.

\vskip 4pt
\paragraph*{\bf Regime of validity.}
The quadratic mixing is resummed exactly, so $\la$ is unrestricted from this effect, while the cubic couplings are treated perturbatively, which requires their dimensionless amplitudes to be small.
Rewriting the couplings in terms of their strong coupling scales, we have the following restrictions.
\begin{itemize}
    \item  $\lambda_2 \dot{\pic}^3$ is a dimension-six coupling so we ask that $(H/\Lambda_\star)\ll 1$ with $\Lambda_\star= \lambda_2^{-1/2}$.
    \item $H \lambda /(2f_{\pi}^2 )\, \sigma \left[\dot{\pic}^2-(\partial_i \pi)^2/a^2\right]$ is a dimension-five coupling so we ask that $H/\Lambda_\lambda \ll 1$ with $\Lambda_\lambda= 2f_\pi^2/( H \lambda )$, i.e. $\lambda \ll \Delta_{\zeta,0}^{-1} \sim \sqrt{\epsilon} M_{\rm Pl}/H$.    At strong mixing, a more restrictive condition is that the highest energy state in the theory, that is $m_{\rm eff}\sim H \lambda$, sits below the strong coupling scale $\Lambda_\lambda$, giving the more stringent bound  $\lambda \ll \Delta_{\zeta,0}^{-1/2}$.
\end{itemize}
Although only the dressed amplitude containing mixing with the isocurvature fluctuation is constrained at CMB scales from observations, $\Delta_{\zeta}=A_s^{1/2}\sim 10^{-4}$, we prove below that the decoupled amplitude $\Delta_{\zeta,0}$ is even smaller so $\Delta_{\zeta,0}^{-1/2} \gtrsim 10^2$, which gives a comfortable range of $\lambda \gg 1$ values for strong quadratic mixing analysis while remaining perturbative at the non-linear level.

In this regime, one could also worry that the cubic interactions $\propto \lambda$ dominate over the ones $\propto \lambda_2^{-1/2}$.
That is a fair point that we treat in great detail in the companion paper~\cite{TBP}, where all discussions related to $\sigma$ modes are included.

Another interesting aspect is whether the strong mixing enhancement found below propagates into loops. This is a question for the resummed loop expansion, controlled by the same small cubic couplings but potentially enhanced fluctuations, which we leave for future work.

Finally, we recall that we restrict ourselves to scale-invariant scenarios (the Wilson coefficients are constant, and we neglect slow-roll corrections) and unit speeds of sound for both curvature and isocurvature fluctuations.

\vskip 4pt
\paragraph*{\bf Integral representation of linear solutions.}
The linear theory defined by Eqs.~\eqref{eq:H2}--\eqref{eq:meff} was recently solved exactly to all orders in $\rho$ in Ref.~\cite{Huenupi:2026abj}, where the mixed mode functions are generated by the action of a non-local operator resumming the confluent structure of the coupled system. 
For non-linear applications, we find it convenient to trade this operator representation for a single integral one, derived in App.~A: each field is expanded over two oscillators $\alpha=1,2$ each of them being a sum of two modes $i=1,2$, and by summing over $i$ every mode collapses onto \emph{channel modes} carrying a single dressed plane wave. 
For $\pi$, writing
\begin{equation}
    \hat{\pic}^{\vec{k}}(\tau) = \pic^\alpha(k,\tau) \hat{a}_\alpha^{\vec{k}} + \text{h. c.}\,,
\end{equation}
we find $\pic^\alpha=\tfrac{H}{\sqrt{2k^3}}A^\alpha_a \Ma^\pi_a(x)$ with $x\equiv-k\tau$ and
\begin{equation}
\label{eq:Mpi}
   \Ma^\pi_a(x)=i\int_0^\infty\!\dd u\,\omega_a(u)\,\frac{1-ix(1+2u)}{1+u}\,e^{ix(1+2u)}\,,
\end{equation}
the whole linear theory being encoded in the weight
\begin{align}
\label{eq:omega}
 \omega_a(u)=  \, & \frac{u^{\frac{ia\la}{2}-1}(1+u)^{-\frac{ia\la}{2}}}{\Gamma\left(\frac{ia\la}{2}\right)} \times \\
   & {}_2F_1\!\left(\tfrac12-\n,\tfrac12+\n;1+ia\la;-u\right), \nonumber
\end{align}
with $\omega_{-a}=\omega_a^*$ and $a=\pm$ the two oscillator channels.
Physically, $u$ labels the comoving frequencies $k(1+2u)$ that dress the massless fluctuation due to mixing with the massive one.
The Bunch-Davies initial conditions fix the Bogoliubov-like coefficients $A^\alpha_a$~\cite{Huenupi:2026abj}, which we find to satisfy
\begin{equation}
\label{eq:diag}
   \sum_\alpha A^\alpha_a(A^\alpha_b)^*=\tfrac12\,e^{a\pi\la/2}\,\delta_{ab}\,,
\end{equation}
which diagonalises all equal-time correlators in the channel index $a$ and introduces the Boltzmann-like weights $e^{a\pi\la/2}$ at the origin of the strong-mixing enhancement described below.

At the boundary $x\to0$ the curvature channel mode collapses onto a number, $\Ma^\pi_a(0)=i r_a$, whose closed form~\eqref{eq:ra} is given in App.~A. 
The dimensionless curvature power spectrum follows immediately from Eqs.~\eqref{eq:diag} and \eqref{eq:ra}, $\Delta_\zeta^2=R\,\Delta_{\zeta,0}^2$ with $\Delta_{\zeta,0}^2=H^2/(8\pi^2\epsilon\mpl^2)$ its decoupled value, that $R=\frac12\sum_{a=\pm} e^{a \pi \la/2} |r_a|^2$, i.e.
\begin{align}
\label{eq:R}
   R &=\left|\frac{\Gamma\!\left(\tfrac34-\tfrac{\n}{2}\right)\Gamma\!\left(\tfrac34+\tfrac{\n}{2}\right)}
   {\Gamma\!\left(\tfrac34-\tfrac{\n}{2}+\tfrac{i\la}{2}\right)\Gamma\!\left(\tfrac34+\tfrac{\n}{2}+\tfrac{i\la}{2}\right)}\right|^2\,,
\end{align}
reproducing Ref.~\cite{Huenupi:2026abj} with our new integral representation.
To gain physical insights, we take various limits of this very general expression:
\begin{itemize}
    \item at weak mixing, $\lambda \ll 1$, $$R=1+\frac{\lambda^2}{4}
\left[
\psi^{(1)}\!\left(\frac34-\frac{\nu}{2}\right)
+
\psi^{(1)}\!\left(\frac34+\frac{\nu}{2}\right)
\right]+\mathcal{O}(\lambda^4) \,,$$ with $\nu=\sqrt{9/4-m^2/H^2}$, the first $\lambda^2$-correction exactly matches perturbative calculations~\cite{Chen:2012ge, Chen:2017ryl};
\item at strong mixing, $\lambda \gg 1$, and fixed bare mass $m$, 
$$R=\pi\sqrt{\la}/\Gamma(3/4)^2\left[1+\pi \nu^2/(4 \lambda) + \mathcal{O}(\lambda^{-2})\right] \,,$$
with the leading order, independent of the mass, coinciding exactly with~\cite{An:2017hlx,Pinol:2023oux};
\item at strong mixing and fixed effective mass $m_{\rm eff}$, 
$$R= \frac{e^{\pi\lambda}}
     {2\pi^{2}\lambda}
\left|
\Gamma\!\left(\frac34-\frac{\nu_{\rm eff}}{2}\right)
\Gamma\!\left(\frac34+\frac{\nu_{\rm eff}}{2}\right)
\right|^{2} \left[1 + \mathcal{O}(\lambda^{-1}) \right] \,,$$ a new result highlighting that, bearing cancellations, the enhancement is $e^{\pi \lambda/2}$ \textit{per leg}.
\end{itemize}

The same construction applies to the isocurvature sector, treated systematically in the companion paper~\cite{TBP} and not needed here.

\vskip 4pt
\paragraph*{\bf New exact primordial bispectrum shapes.}
Since the mixing is resummed into the modes, any tree-level bispectrum arises at first order in the interaction~\eqref{eq:H3}, from a \emph{single} vertex whose three legs reach the boundary, and every leg carries a single shifted plane wave.
As shown in App.~B, the in-in time integral of any scale-invariant single-vertex diagram evaluates, via a Schwinger parameter $\xi$ that disentangles the frequencies, to a $\xi$-integral over \emph{independent leg kernels}
\begin{equation}
\label{eq:legkernels}
   \Wn_n^a(\beta)=\int_0^\infty\!\dd u\,\omega_a(u)\,\frac{(1+2u)^n}{1+u}\,e^{-\beta u}\,,
\end{equation}
evaluated at rescaled arguments $\beta_j=2\xi e_j$, with $e_j=k_j/k_t$, $k_t=k_1+k_2+k_3$: a velocity leg calls $\Wn_2$ and an undifferentiated $\pic$ leg $\Wn_0+e_j\xi\,\Wn_1$. 
This reduces \emph{all} scale-invariant tree-level one-vertex diagrams, at any mixing strength, to a single Schwinger-parameter integral over pre-computable kernels.
Related but different integral representations with applications to cosmological correlators were recently devised in~\cite{Werth:2024mjg,Belrhali:2026rkn,Belrhali:2026ygh}.

We now specialise to $\dot{\pi}_{\rm c}^3$. Collecting the boundary factors $ir_{a_j}$ of the external ends, the Boltzmann weights of Eq.~\eqref{eq:diag}, and defining the dimensionless shape function $S(k_1,k_2,k_3)\equiv(k_1k_2k_3)^2B_\zeta/[(2\pi)^4\Delta_\zeta^4]$, we obtain the new exact bispectrum shapes
\begin{equation}
\label{eq:Spi3}
\begin{aligned}
   S=\, &\mathcal{N}e_1e_2e_3 \times \\
&\Real\Big[\sum_{\vec{a}}\Big(\prod_{j=1}^3 e^{a_j\pi\la/2}r_{a_j}\Big)
   \!\!\int_0^\infty\!\!\!\dd\xi\,\xi^2e^{-\xi}\!\prod_{j=1}^3 \Wn_2^{-a_j}(2\xi e_j)\Big],\\
   &\hspace{0.9cm}\mathcal{N}=\frac{3}{32\pi R^{3/2}}\left(\frac{H}{\Lambda_\star}\right)^2\frac{1}{\Delta_\zeta}\,,\qquad \Lambda_\star\equiv\lambda_2^{-1/2}\,,
\end{aligned}
\end{equation}
where $\vec{a}=(a_1,a_2,a_3)$ runs over the $2^3$ channel assignments, $(H/\Lambda_\star)^2\Delta_\zeta^{-1}$ is the usual bispectrum amplitude of the EFT of inflation in terms of its strong-coupling scale $\Lambda_\star$, and the factor $R^{-3/2}$ descends from the measured $\Delta_\zeta^4\propto R^2$ in the denominator of the definition of the shape and rewriting $\Delta_{\zeta,0}^{-1} =  R^{1/2} \Delta_{\zeta}^{-1}  $.
Once the kernels~\eqref{eq:legkernels} are tabulated---the integral is always convergent---evaluating Eq.~\eqref{eq:Spi3} over the whole momentum triangle takes a fraction of a second on a laptop, which constitutes a substantial improvement compared to already known semi-analytical and numerical methods.

To quantify how the resummed mixing deforms the shape, we use the standard cosine correlation of data analyses~\cite{Babich:2004gb},
\begin{equation}
\label{eq:cosine}
   \cos(S,S')=\frac{\braket{S,S'}}{\sqrt{\braket{S,S}\braket{S',S'}}}\,,
\end{equation}
where $\braket{S,S'}$ integrates $SS'$ with uniform weight over the triangle configurations $(x_1,x_2)=(k_1/k_3,k_2/k_3)$, $x_1\leq x_2\leq 1$ and $x_1+x_2\geqslant1$, sampling $x_1\geqslant10^{-3}$. The right panel of Fig.~\ref{fig:shape_cosmap} shows $\cos(S,S_{\rm eq})$ with the equilateral shape $S_{\rm eq}\propto e_1e_2e_3$ for effectively heavy fields, across the $(\la,\muf)$ plane. At weak mixing the no-exchange bispectrum is indistinguishable from the equilateral shape, $\cos(S,S_{\rm eq})>0.99$ for $\la\lesssim 0.5$. At strong mixing, instead, the shape decorrelates in a whole band of the $(\la,\muf)$ plane, with $\cos(S,S_{\rm eq})$ crossing zero (dashed lines) and reaching $\simeq -0.9$.
As shown in the left panel, these new shapes do not merely correspond to anti-equilateral nor orthogonal shapes; they are invaded, already at mild squeezing, by the collider oscillations of the effectively heavy isocurvature field, a purely multifield feature which is usually only visible in the deep squeezed limit.
Remarkably, the decorrelation band runs along the bare $m^2=0$ dotted line $\la=\sqrt{9/4+\muf^2}$; below the band, increasing $\muf$ at fixed $\la$ Boltzmann-suppresses the oscillations and restores the equilateral correlation with $\cos(S,S_{\rm eq})>0.99$, while above it, in the bare-tachyonic regime, increasing $\lambda$ at fixed $\muf$ pushes back the clock signal to very squeezed regions, partially restoring equilateral correlation until it degrades again, e.g. $\cos(S,S_{\rm eq})\simeq0.9$ in the upper left corner.

\subparagraph{The squeezed limit.}
In the squeezed limit $\ka\equiv k_1/k_3\ll 1$, the soft leg's argument becomes $\beta=\kappa \xi \ll 1$.
Indeed, the $\xi$-integral contains $e^{-\xi}$, effectively cutting $\xi \gtrsim 1$ contributions.
Now, for small $\beta$, Eq.\eqref{eq:legkernels} is dominated by the large-$u$ tail of its integrand, since the effective domain of integration becomes of size $\sim 1/\beta\gg 1$.
We can therefore expand $\omega_a(u)$ and the rational function of $u$ of the soft leg as $u\gg 1$, with ${}_2F_1 \sim u^{-1/2\pm \nu_{\rm eff}}$ so that $\Wn_n^a$'s integrand scales as  $u^{n-3/2\pm \nu_{\rm eff}} $.
Keeping track of all factors and performing the Laplace transform explicitly, we get at leading order $\Wn_2^a(\beta) \underset{\beta \ll 1}{\simeq}\sum_{b=\pm}4\,\mathcal{W}^a_{b}\,\beta^{-\frac12-b\n}$,
\begin{equation}
\label{eq:tails}
    \text{with} \quad 
   \mathcal{W}^a_{\pm}=\frac{\Gamma(1+ai\la)\,\Gamma(\pm2\n)}{\Gamma\!\left(\frac{ai\la}{2}\right)\Gamma\!\left(\frac12\pm\n+ai\la\right)}\,.
\end{equation}
The non-analyticity in $\ka$ of the squeezed bispectrum follows:
\begin{align}
\label{eq:Csqueeze}
S\underset{\ \kappa \ll 1}{\simeq} & \mathcal{N}\times \Real\!\Big[\sum_{b=\pm}\mathcal{C}_b\,\ka^{\frac12-b\n}\Big]\,,\\
   \mathcal{C}_\pm=&\,\tfrac12\sum_{\vec{a}}\Big(\prod_j e^{a_j\pi\la/2}r_{a_j}\Big)
   \mathcal{W}_{\pm}^{-a_1}J_{\pm}^{-a_2,-a_3}\,, \nonumber
\end{align}
with $J_{\pm}^{b_2b_3}=\int_0^\infty\dd\xi\,\xi^{\frac32\mp\n}e^{-\xi}\Wn_2^{b_2}(\xi)\Wn_2^{b_3}(\xi)$ pre-computable hard integrals: the squeezed limit is known in a closed form at any $\la$.
For a heavy field, $\n=i\muf$, only the combination $\mathcal{C}_++\mathcal{C}_-^*$ is observable, and the two branches combine into the collider clock
\begin{equation}
\label{eq:clock}
   S\underset{\ \kappa \ll 1}{\simeq}  \mathcal{N} \sqrt{\ka}\;\mathcal{A}(\muf,\la)\,\sin\!\left[\muf\ln\ka-\delta(\muf,\la)\right]\,,
\end{equation}
with $\mathcal{A}e^{i\delta}=-i(\mathcal{C}_++\mathcal{C}_-^*)$, while a light field yields a pure power law $\ka^{\frac12-\n}$.

Three properties are noteworthy. First, the frequency is \emph{exactly} $\muf$ at every mixing strength as evidenced in previous semi-analytical and numerical works~\cite{An:2017hlx,Werth:2023pfl,Pinol:2023oux}.
Second, a collider signal is present already with cubic interactions involving only $\pi$: the quadratic mixing suffices to create the massive-field oscillations.
Third, the channel weights $e^{a\pi\la/2}$ in Eq.~\eqref{eq:Csqueeze} act as a non-perturbative Boltzmann asymmetry between the two dressed branches: at strong mixing they exponentially enhance the clock amplitude relative to its perturbative $\la^2$ scaling, and rotate its phase $\delta$ by several radians, analogously to the chemical-potential mechanism~\cite{Bodas:2020yho,Sou:2021juh} but without invoking additional physics. 
Fig.~\ref{fig:clock} displays these features: the left panels compare the exact shape with the closed form~\eqref{eq:clock} and its weak mixing asymptotics (see below) and the right panel shows the complex amplitude $\mathcal{A} e^{i \delta}/\la^2$, interpolating between the perturbative regime and the strongly enhanced, de-phased one. 

As $\la\ll 1$ the weight localises, $\omega_a(u)\to\delta(u)$ in the sense of distributions, so $\Wn^a_n\to1$, $r_a\to1$, and Eq.~\eqref{eq:Spi3} collapses at fixed configuration to the standard single-field result $B_\zeta\propto1/(k_1k_2k_3k_t^3)$~\cite{Cheung:2007st,Senatore:2009gt}, of equilateral type and clockless. The collider signal appears at next order: the $\O(\la)$ corrections to the kernels and the leading tail~\eqref{eq:tails} are \emph{odd} in $a$, so the channel sum in Eq.~\eqref{eq:Csqueeze} cancels at $\O(\la)$ and the clock coefficient starts at $\O(\la^2)$,
\begin{align}
\label{eq:Cweak}
   \mathcal{C}_+&\underset{\lambda \ll 1}{\simeq}\,\frac{\Gamma(2\n)\Gamma\!\left(\tfrac52-\n\right)}{\Gamma\!\left(\tfrac12+\n\right)}\,\mathcal{S}(\n)\,\la^2+\O(\la^4)\,,\nonumber \\
   \mathcal{S}(\n)&=\psi\!\left(\tfrac12+\n\right)-\frac{\psi\!\left(\tfrac34-\tfrac{\n}{2}\right)+\psi\!\left(\tfrac34+\tfrac{\n}{2}\right)}{2} \nonumber\\
   & \hspace{0.4cm}-\ln2-\frac{i\pi}{2}\,,
\end{align}
with $\mathcal{C}_-=\mathcal{C}_+(\n\to-\n)$. 
This scaling matches the minimal perturbative diagram---two insertions of the mixing on the soft leg---and Eq.~\eqref{eq:Cweak} must reproduce the corresponding perturbative calculation for the amplitude and phase, which is not available in the literature to our knowledge.
In Fig.~\ref{fig:clock}, we find that the exact amplitudes and phases indeed approach these asymptotics as $\la\ll1$, a consistency check of the whole framework.

The strong-mixing growth at fixed effective mass, visible in Fig.~\ref{fig:clock}, can be analytically derived.
At leading order, each of the three dressed external leg contributes $e^{\pi\la/2}|r_+|\sim \lambda^{-1/2} e^{3\pi\la/4}$, the soft-leg tail grows as $|\mathcal{W}^{-}_\pm|\sim \lambda e^{\pi\la/4}$ and the hard integrals as $|J_\pm^{-,-}|\sim \lambda^{1/2} e^{\pi\la/2}$, so that each branch coefficient grows as $|\mathcal{C}_\pm|\sim e^{3\pi\la}$ (matching the expected scaling of $e^{\pi \lambda /2}$ per leg derived below Eq.~\eqref{eq:R} in the strong mixing regime, times three external legs and three internal ones). 
However, this leading asymptotics is either purely imaginary for effectively light fluctuations or such that $\mathcal{C}_+ = - \mathcal{C}_-^* $ for heavy ones, so in both cases it cancels in the final amplitude, see Eqs.~\eqref{eq:Csqueeze}--\eqref{eq:clock}.
The leading non-vanishing order, therefore, imposes considering one $a_j = -1$ permutation, with $e^{-\pi\la/2}|r_-|\sim \lambda^{-1/2} e^{-\pi\la/4}$ in one external leg, the rest being unaffected, resulting in a total growth $\sim e^{2\pi\la}$.
Since $\mathcal{N}\propto R^{-3/2}\propto\la^{3/2}e^{-3\pi\la/2}$ at fixed $\nu_{\rm eff}$, the physical amplitude in the squeezed limit grows as 
\begin{equation}
\label{eq:fNL}
    f_{\rm NL}^{\rm squeezed} \underset{\lambda \gg 1}{\simeq} 
    \left(\frac{H}{\Lambda_\star}\right)^2 \frac{\lambda^{3/2}
    e^{\pi\la/2}}{\Delta_\zeta} g(\nu_{\rm eff}) \,,
\end{equation}
with $g(\nu_{\rm eff}) $ a decreasing function of the effective mass only, and multiplying the scaling $\kappa^{1/2-\nu_{\rm eff}}$---a power law for effectively light fluctuations or the clock signal for heavy ones.
Although $(H/\Lambda_\star)^2 \ll 1$ by perturbativity, the amplitude is boosted by $A_s^{-1/2}\sim 10^{4-5}$ and by the strong mixing $\lambda \gg 1$, making for large squeezed signals.

\vskip 4pt
\paragraph*{\bf Conclusion.}
In this \emph{Letter}, we have presented the first exact analytical computation of a primordial bispectrum with the quadratic mixing between curvature and isocurvature fluctuations resummed to all orders. The key structural result---every tree-level bispectrum diagram reduces to a single Schwinger-parameter integral over pre-computable leg kernels---turns the strongly mixed regime, previously the preserve of numerics, into an analytical laboratory. Applied to the minimal $\dot{\pic}^3$ interaction, we revealed a phenomenology that interpolates between the single-field equilateral shape at weak mixing and genuinely new multifield shapes at strong mixing.
We unveiled an exponentially enhanced squeezed signal with kinematical dependence set by $\nu_{\rm eff}$, and we obtained its closed form at any $\la$.
Since these shapes do not correlate with the equilateral template and do not resemble the orthogonal one, our results motivate a dedicated data-analysis effort~\cite{Planck:2019kim,Meerburg:2016zdz,Sohn:2024xzd,Philcox:2025bbo,Suman:2025vuf,Kumar:2026dih}, for which the efficiency of our representation~\eqref{eq:Spi3} will be instrumental.

The strong-mixing regime at fixed $\nu_{\rm eff}$ eventually explores a tachyonic bare mass, $m^2<0$ for $\la^2>9/4-\nu_{\rm eff}^2$.
But this instability is not dramatic.
Indeed, it only operates on sub-Hubble scales for a limited amount of time for each $k$ mode, all fluctuations converging outside the Hubble radius where the positive $\meff^2$ takes over---the mechanism at play in sidetracked inflation, in EFTs with an imaginary speed of sound, and in hyperinflation~\cite{Garcia-Saenz:2018ifx,Garcia-Saenz:2018vqf,Brown:2017osf,Aoki:2026qea}, and distinct from the tachyon collider of Ref.~\cite{McCulloch:2024hiz} for which $\meff^2$ itself is negative.

In the companion paper~\cite{TBP}, we develop the complete diagrammatic rules for the resummed theory, apply them to diagrams with cubic interactions involving isocurvature fluctuations (the would-be exchange diagrams of the perturbative-in-$\lambda$ approach).
Theoretically, a natural extension would be the study of strongly mixed spinning fluctuations, for which the soft limit collider signal depends on the angle from which it is approached~\cite{Bordin:2018pca}.
Phenomenologically, a more detailed study of effectively light fluctuations is also required.
Finally, our new exact bispectrum shapes in multifield inflation motivate a dedicated CMB data analysis, as well as forecasts for upcoming galaxy surveys.

\subparagraph*{Note added.} As this article was close to completion, I became aware of a forthcoming work~\cite{WIP}, by the authors of Ref.~\cite{Huenupi:2026abj}, where a similar integral representation is used, and focus is placed on the squeezed limit of the bispectrum, though from different cubic interactions.

\begin{acknowledgments}
I am grateful to 
Matteo Braglia, 
Sebastian Cespedes,
and the authors of Ref.~\cite{Huenupi:2026abj}
for useful discussions related to this work.
\end{acknowledgments}

\bibliography{biblio}

\subfile{appendix-letter}

\end{document}

%% file: appendix-letter.tex
\section*{Supplemental Material}
\label{appendix}

\paragraph*{\bf Appendix A: Integral representation of the exact linear solutions.}
Each field is expanded on two oscillators, $X_{\vec{k}}(\tau)=X^\alpha(\tau,k)\hat{a}^{\vec{k}}_\alpha+{\rm h.c.}$ with $\alpha=1,2$ and $X\in\{\pic,\sigma\}$, and each mode function on the two Bunch--Davies carriers of its decoupled dynamics, e.g.
\begin{equation}
\label{eq:carriers}
\pic^\alpha(\tau,k)=\frac{H}{\sqrt{2k^3}}\,\pi^\alpha_i(x)\,u_\pi^i(x)\,,\quad u^1_\pi=i(1-ix)e^{ix}\,,
\end{equation}
with $u^2_\pi=(u^1_\pi)^*$ and $x\equiv-k\tau$, and similarly for $\sigma$ in terms of $u_\sigma^i$ containing the usual Hankel functions but with parameter $\nu_{\rm eff}$. Reference~\cite{Huenupi:2026abj} solved the coupled system, rewritten as a fourth-order linear equation, as an infinite series of derivatives of Tricomi confluent functions $U$.
By inserting the standard integral representation $U(a,b,x)=\Gamma(a)^{-1}\int_0^\infty\dd u\, u^{a-1}(1+u)^{b-a-1}e^{-x u}$,
derivatives with respect to $x$ simply become multiplications by $-u$ and the series representation can be resummed into a ${}_2F_1$ with argument $-u$, at the price of keeping the $u$-integral.
Re-organising, we get
\begin{equation}
\label{eq:ansatz}
   \pi_1^\alpha=A^\alpha_a f_1^a\,,\qquad f_1^a(x)=\int_0^\infty\!\dd u\,\omega_a(u)\,e^{2iux}\,,
\end{equation}
with $a=\pm$ summed over, $\omega_a(u)$ given by Eq.~\eqref{eq:omega}, together with $\pi_2^\alpha=A^\alpha_a f_2^a$ and $f_2^a(x)=\int_0^\infty\dd u\,\omega_a(u)\,\tfrac{u}{1+u}\,e^{2i(1+u)x}$.
Matching the Bunch--Davies behaviour at $x\to\infty$ fixes the Bogoliubov-like coefficients to $A^1_\pm=\tfrac12 e^{\pm\pi\la/4}$, $A^2_\pm=\pm\tfrac{i}{2}e^{\pm\pi\la/4}$~\cite{Huenupi:2026abj}, which yields Eq.~\eqref{eq:diag}.

The virtue of this representation appears upon summing over the carriers $i\in\{1,2\}$ explicitly. Since $u_\pi^1(x)\,e^{2iux}$ and $u_\pi^2(x)\,e^{2i(1+u)x}$ carry the very same phase $e^{ix(1+2u)}$, each oscillator channel $a=\pm$ collapses onto the single dressed plane wave $M_a^\pi(x)$ of Eq.~\eqref{eq:Mpi}.

At the boundary, we get that $M_a^\pi(0) = i r_a$ with
\begin{equation}
\label{eq:ra}
   r_a
   =\frac{\Gamma\!\left(\tfrac12+\tfrac{ai\la}{2}\right)\Gamma\!\left(\tfrac34-\tfrac{\n}{2}\right)\Gamma\!\left(\tfrac34+\tfrac{\n}{2}\right)}
   {\sqrt{\pi}\,\Gamma\!\left(\tfrac34-\tfrac{\n}{2}+\tfrac{ai\la}{2}\right)\Gamma\!\left(\tfrac34+\tfrac{\n}{2}+\tfrac{ai\la}{2}\right)}\,,
\end{equation}
and such that $r_-=r_+^*$. 
The closed form is obtained under the change of $ u \rightarrow  t/(1 - t)$ that renders the $t$-integrand well behaved and analytically integrable for $ t\in [0,1]$, giving Eq.~\eqref{eq:ra}. 
We have verified that our new integral representation reproduces exactly the linear solutions of Ref.~\cite{Huenupi:2026abj}, thereby setting a robust stage for the non-linear applications of this work.

\vskip 6pt
\paragraph*{\bf Appendix B: Reduction to a single Schwinger integral.}
In the in-in formula, the bispectrum $B_\pic\propto\int\dd\tau\,\braket{\hat{\pic}^{\vec{k}_1}\hat{\pic}^{\vec{k}_2}\hat{\pic}^{\vec{k}_3}\,\hat{\mathcal{H}}^{\rm int}(\tau)}$, each dressed leg~\eqref{eq:Mpi} carries a single shifted plane wave, so the time integrand is $\propto\tau^{N}\prod_j\int\dd u_j\,\omega_{a_j}(u_j)(\dots)e^{-ik_j\tau(1+2u_j)}$ against the vertex measure $a(\tau)\dd\tau=-\dd\tau/(H\tau)$, with $N$ set by the interaction. Writing $\mathcal{E}=1+2\sum_je_ju_j$, the time integral evaluates to
\begin{equation}
\label{eq:schwinger}
\begin{aligned}
   \int_{-\infty}^{0}\!\dd\tau\,(-\tau)^{N}\,e^{ik_t\tau\mathcal{E}}
   &=\frac{N!}{(ik_t\mathcal{E})^{N+1}}\\
   =\frac{1}{(ik_t)^{N+1}}\int_0^\infty\!\dd\xi\,&\xi^{N}e^{-\xi}\prod_j e^{-2\xi e_j u_j}\,,
\end{aligned}
\end{equation}
with the Bunch--Davies rotation left implicit: the Schwinger parameter $\xi$ disentangles the frequencies and the $u_j$ integrals factorise into the leg kernels~\eqref{eq:legkernels} at $\beta_j=2\xi e_j$.
It is immediately clear that this proof applies to any scale-invariant tree-level contact diagram with any number of external legs.
It also applies to interaction Hamiltonians that include the $\hat{\sigma}$ operator, since, due to the non-perturbative treatment of the quadratic mixing, $\left[\hat{\pic},\hat{\sigma}\right]\neq 0$ even for interaction picture fields. Therefore, the usual ``exchange" bispectrum diagrams in multifield inflation become contact ones, and the rest of the proof proceeds.